# Revisiting apparent ideal diamagnetism at ambient conditions in graphene–n-heptane–permalloy systems

Rajendra Dulal[†], Serafim Teknowijoyo[†], Sara Chahid[†], Vahan Nikoghosyan[*,†], and Armen Gulian[†,‡]

[†]*Laboratory of Advanced Quantum Materials and Devices, Institute for Quantum Studies, Chapman University, Orange, CA 92866, USA*
[*]*Institute for Physics Research, National Academy of Sciences, Ashtarak-2, 0304 Armenia*
[‡]*gulian@chapman.edu*

We previously reported apparent ideal diamagnetism at ambient conditions in a graphene–n-heptane–permalloy system. At the same time, the experiments revealed inconsistent behavior, including signal freezing and occasional paramagnetic responses. Further measurements performed without graphene produced similar signals, indicating that graphene is not responsible for the observed effects. The results suggest that magnetic field redistribution caused by inhomogeneities in the permalloy foil and experimental geometry can mimic ideal diamagnetism in sub-milligauss measurements. These findings revise the interpretation of our earlier results and emphasize caution in interpreting ultra-low-field magnetic measurements.

In 2020 we published an article [1] which described a very non-trivial magnetic response of graphene when immersed into n-heptane in presence of permalloy foil. Experiments described in the article [1] were principally similar to the report described in the preprint [2]. For performing experiments in close to ideal conditions we used setup illustrated in Fig. 1.

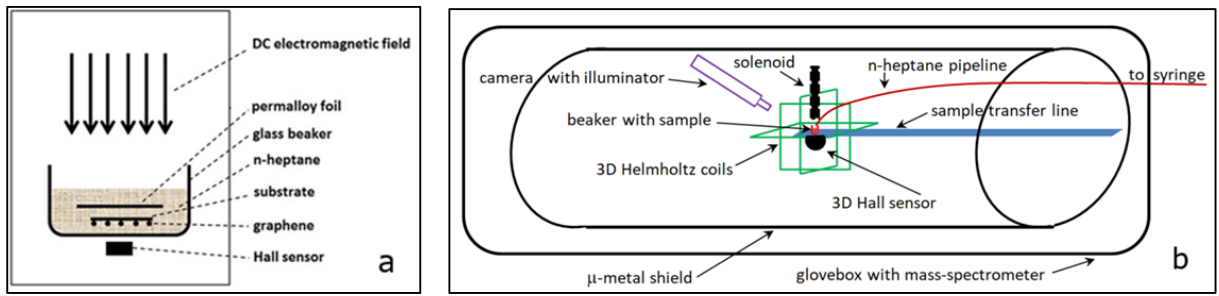

**Fig. 1**. (a) Schematic illustration of the experimental concept: graphene on a substrate covered by a thin permalloy foil appears to screen the applied magnetic field after n-heptane is injected into the beaker.
(b) Experimental setup inside a magnetically shielded enclosure. The external magnetic field at the sample location was reduced using multilayer permalloy shielding and three-axis Helmholtz coils. Hall sensors were used to monitor the magnetic field with sensitivity down to 0.01 mG in the absence of the sample.

Because the previously reported effect suggested possible room-temperature superconductivity, clarification of these observations is important.

The experiments were performed in the following sequence. Single- or double-layer graphene samples on Si/SiO2 substrates were first examined using Raman-mapping microscope, then they were placed into a glass beaker with graphene layer down, and on the top was placed a flat permalloy foil. This initially dry sample system (Fig. 1a) was inserted into the specially prepared chamber using the mechanically driven transfer line (see Fig. 1b), which was delivering the sample into the active area. The active area was protected from the Earth's magnetic field by multilayer permalloy tube with high aspect ratio. The remnant field was additionally diminished by a 3D Helmholtz coil, which allowed regulation of all three components of the magnetic field vector using the feedback from the Hall sensors located under the sample location plane, very close to the sample. Figure 1b also shows the n-heptane pipeline, which also was mechanically driven: at the arrival of sample it was elevated to allow beaker to occupy its state precisely, and then in was lowered to deliver the liquid at the next stage. The other end of the pipe was connected to a syringe, which was hand activated to deliver the proper dosage of the n-heptane to the beaker. All the events inside of the µ-metal tube were possible to visualize with a zooming camera. Prior to delivering the liquid, the magnetic field was being applied via the electromagnet; it was made sure that the applied field is stably affecting the system. All three axes (x, y, and z) data delivered by Hall sensors were being recorded (the main signal corresponded to vertical z-direction) using LabVIEW and Keithley-181 nanovoltmeters. Digital files also contained the initial and final markers of liquid injection.

---
[‡] Corresponding author

The results which were pointing towards the ideal diamagnetic effect are summarized in the Figs. 2-4. Figure 2 demonstrates the case allowing direct interpretation as an ideal diamagnetism in graphene-n-heptane-permalloy system (original curves are in Ref. 1).

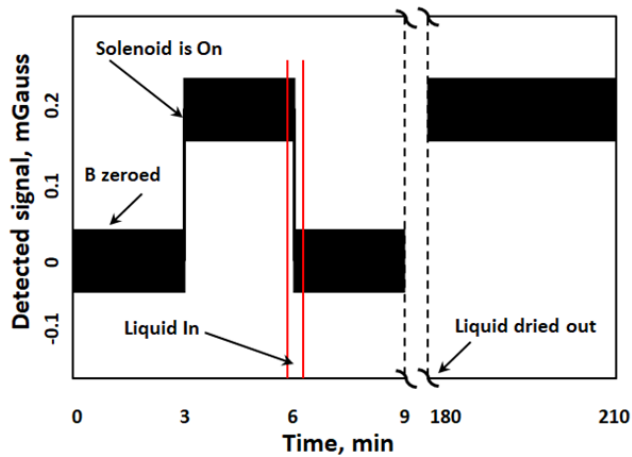

**Fig. 2**. Schematic representation of apparent ideal diamagnetic behavior reported in Ref. 1. Injection of n-heptane reduces the detected magnetic signal (~0.2 mG) to near zero. The signal gradually recovers after evaporation of n-heptane (approximately 3 hours).

This kind of behavior has not been reproducible, and it took very many efforts to achieve it. To find out the reasons of unsuccessful runs, we were closely monitoring conditions of all experiments, using very dry atmosphere to eliminate the negative contribution of water vapors, examining the behavior in argon and nitrogen atmospheres. Graphene samples from different sources have been tested. During these experiments we also repeatedly obtained somewhat strange behavior of the system (Fig. 3).

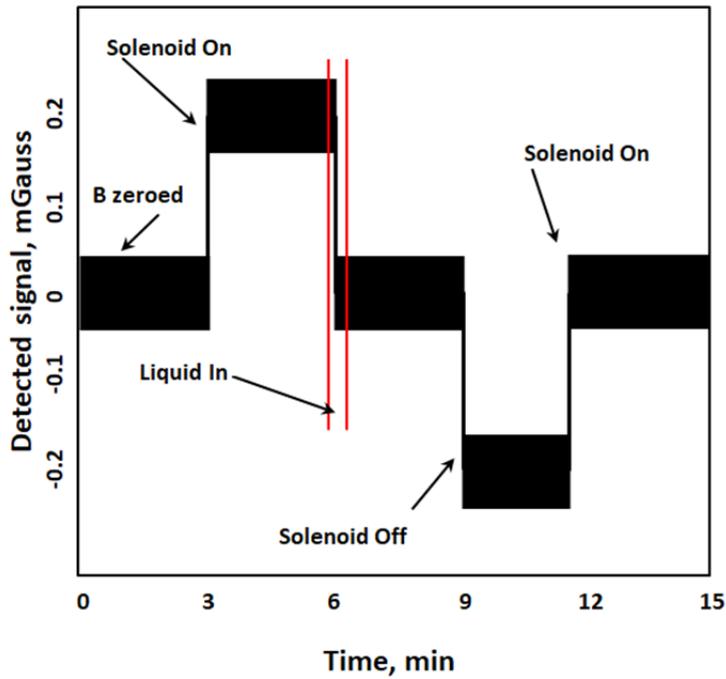

**Fig. 3**. Schematic representation of the "freezing" effect observed in Ref. 1. After injection of n-heptane, the compensating signal persists even after switching off the external magnetic field, resulting in a residual signal of approximately −0.2 mG. Reapplying the external field restores the initial condition.

Such a behavior contradicts the Meissner-type interpretation of the effect. Further difficulties arise with the paramagnetic response, which sometimes has been detected (Fig. 4).

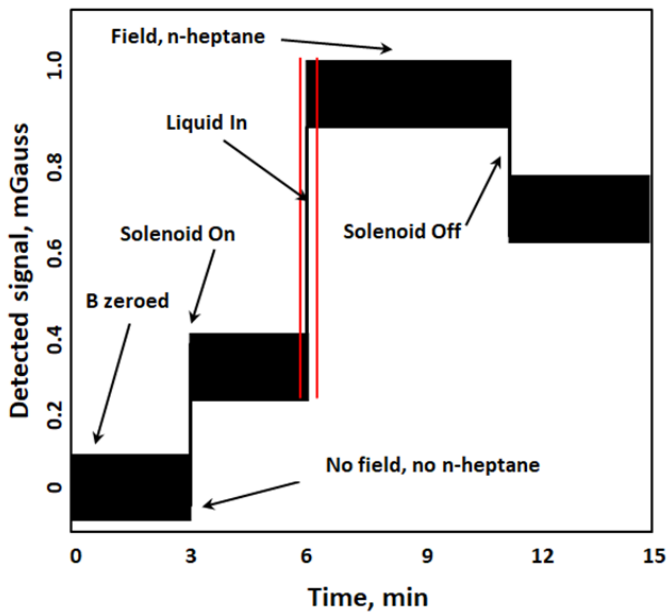

**Fig. 4**. Schematic representation of an "antidiamagnetic" (paramagnetic - like) response observed after n-heptane injection in Ref. 1.

Because of these puzzling data, we concluded our article [1] with a statement that more experimental exploration is required, and until then, we refrain from physical interpretation of the observed facts.

Later, an additional experimental observation was made (Fig. 5), which we think is worth to discuss.

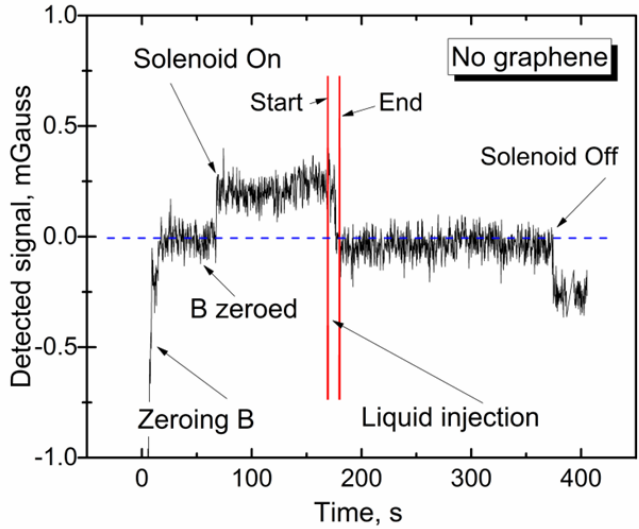

**Fig. 5**. Magnetic response measured in the absence of graphene. After zeroing the background magnetic field, an external field was applied. Injection of n-heptane resulted in compensation of the magnetic signal, which remained "frozen" after switching off the external magnetic field.

This observation sheds light on the effects discussed above. The close analogy with previously obtained results (reported in [1] and schematically represented in Figs. 2, 3 and 4) indicates that graphene does not contribute to the observed signals. The possibility that the observed behavior originates from a permalloy–n-heptane system alone is also unlikely, particularly given the occasional paramagnetic ("antidiamagnetic") response shown in Fig. 4. A plausible explanation is that inhomogeneities in the permalloy foil lead to redistribution of magnetic field lines. The injection of n-heptane plays a role of micro-movement of the permalloy foil, causing the observable peculiarities. Drying of the liquid also can cause micromotion which may help to understand the results in Fig. 1.

Indeed, in 1973 Mallinson discovered a "magnetic curiosity": when a planar structure, such as a flat plate, tape, or disc, possesses in-plane and out-of-plane components of magnetization varying as $\mu_0\sin(kx)$ and $\mu_0\cos(kx)$, respectively, all the external fringing flux emerges on one side of the structure, while the magnetic field on the opposite side is suppressed [3,4].

Figure 6 illustrates this effect using COMSOL modeling of a thin magnetic layer with an imposed rotating magnetization pattern. The calculation is intended as a qualitative demonstration of magnetic-field redistribution rather than a quantitative model of the soft permalloy foil used in the experiments. In a real permalloy foil, nonuniform induced magnetization and small mechanical displacements may produce similar redistribution of magnetic field lines.

In such configurations, the local magnetic field at the Hall sensor position may be reduced or even effectively canceled, thereby mimicking an ideal diamagnetic response. At the same time, the ferromagnetic nature of permalloy may lead to enhanced signals in the opposite direction, which is consistent with the occasional paramagnetic responses observed experimentally.

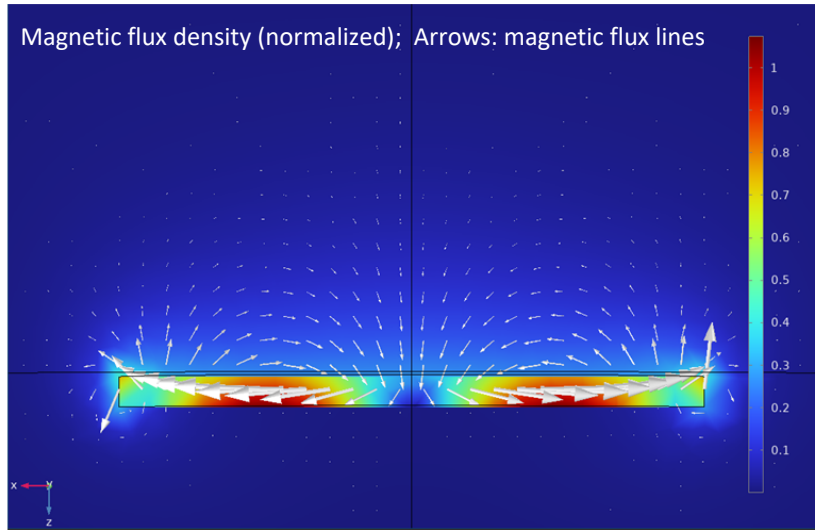

**Fig. 6**. Illustration of one-sided magnetic flux (Mallinson effect) obtained using COMSOL modeling of a thin magnetic layer with an imposed rotating magnetization pattern. In such configurations, magnetic flux emerges predominantly on one side of the structure, while the field on the opposite side is strongly suppressed. This calculation [5] is intended as a qualitative demonstration of magnetic-field redistribution rather than a quantitative model of the soft permalloy foil used in the experiments.

In summary, this report is not intended to dismiss recent claims of possible superconductivity-related features reported in [6]. Rather, we emphasize that our own experimental findings do not support the presence of ideal diamagnetism in the graphene–n-heptane–permalloy system. These observations clarify the interpretation of our earlier results and highlight the importance of careful control of magnetic field distribution in ultra-low-field measurements.

Acknowledgments

This work was supported in part by the ONR Grants N00014-16-1-2269, N00014-17-1-2972, N00014-18-1-2636, and N00014-19-1-2265. We are grateful to Dr. Kawashima for multiple discussions on the related topics and his continued support of our research.

# References

1. Y. Kawashima, R. Dulal, S. Teknowijoyo, S. Chahid, and A. Gulian. Ideal diamagnetic response at room temperature by graphene-n-heptane-permalloy system. *Mod. Phys. Lett.* B **34**, 2050415 (2020).
2. Y. Kawashima, Observation of the Meissner effect at room temperature in single-layer graphene brought into contact with alkanes. *arXiv: 1801.09376* (2018).
3. J.C. Mallinson, One-sided fluxes – A magnetic curiosity?, *IEEE Trnas. Magn.* **MAG-9**, 678 (1973).
4. H.A. Shute, J.C. Mallinson, D.T. Wilton, and D. J. Mapps, One-sided fluxes in planar, cylindrical, and spherical magnetized structures, *IEEE Trnas. Magn.* **36**, 440 (2000).
5. We found the stationary solution for a foil of radius 10mm and thickness 1mm using equations: div**B**=0, **B**=$\mu_0\mu_r$**H**+**B**$_r$, **B**$_f$=|**B**$_r$|**e**/|**e**|, **H**=−**grad**$V_m$, where **e**={sin(kx),0,cos(kx)}, $V_m$ is magnetic scalar potential, and boundary conditions **n**·**B**=0 are applied (more details can be found in COMSOL 6.4 Application Libraries, AC/DC Module, Introductory Magnetostatics, one_sided_magnet).
6. Y. Kawashima, Superconducting characteristics of a graphite/n-alkane mixture above room temperature, *Res. Square* (2024) DOI: https://doi.org/10.21203/rs.3.rs-4851080/v1